\documentclass[final,5p,times,twocolumn]{elsarticle}

\usepackage{amssymb}
\usepackage{amsmath}
\usepackage{svg}
\journal{Applied Radiation and Isotopes}
\begin{document}

\begin{frontmatter}

%% Title, authors and addresses

%% use the tnoteref command within \title for footnotes;
%% use the tnotetext command for theassociated footnote;
%% use the fnref command within \author or \affiliation for footnotes;
%% use the fntext command for theassociated footnote;
%% use the corref command within \author for corresponding author footnotes;
%% use the cortext command for theassociated footnote;
%% use the ead command for the email address,
%% and the form \ead[url] for the home page:
%% \title{Title\tnoteref{label1}}
%% \tnotetext[label1]{}
%% \author{Name\corref{cor1}\fnref{label2}}
%% \ead{email address}
%% \ead[url]{home page}
%% \fntext[label2]{}
%% \cortext[cor1]{}
%% \affiliation{organization={},
%%             addressline={},
%%             city={},
%%             postcode={},
%%             state={},
%%             country={}}
%% \fntext[label3]{}

\title{First Pilot Tests of Compton Imaging and Boron Concentration Measurements in BNCT Using i-TED}
%\title{Real-Time Boron Concentration Measurement in BNCT Using Compton Imaging}

%% use optional labels to link authors explicitly to addresses:
 \author[1]{J.~Lerendegui-Marco\footnote{jorge.lerendegui@ific.uv.es}}

\author[1]{J.~Balibrea-Correa}

\author[2,3]{P.~\'Alvarez-Rodr\'iguez}

\author[1]{V.~Babiano-Su\'arez}

\author[1]{B.~Gameiro}

\author[1]{I.~Ladarescu}

\author[3,4]{C.~Méndez-Malagón} 

\author[2]{C. Michelagnoli}

\author[5]{I. Porras}

\author[3]{M. Porras-Quesada}

\author[4]{C. Ruiz-Ruiz}

\author[1,5]{P. Torres-S\'anchez}

\author[1]{C.~Domingo-Pardo}

\affiliation[1]{organization={Instituto de Física Corpuscular, CSIC-Universitat de València},country={Spain}}
\affiliation[2]{organization={Institut Laue-Langevin},country={France}}
\affiliation[3]{organization={Instituto de Biopatología y Medicina Regenerativa, Centro de Investigación Biomédica. Universidad de Granada},country={Spain}}
\affiliation[4]{organization={Departamento de Bioquímica y Biología Molecular III e Inmunología, Universidad de Granada},country={Spain}}
\affiliation[5]{organization={Departamento de Física Atómica, Molecular y Nuclear, Universidad de Granada},country={Spain}}

%% Abstract
\begin{abstract}
%% Text of abstract
Dosimetry in BNCT poses significant challenges due to the indirect effect of neutrons interacting with elements within the body and uncertainties associated with the uptake of boron compounds used in clinical practice. Current treatment planning relies on unconventional estimates of boron tumor uptake derived from prior PET scans and thus, an online boron-uptake monitor would be highly convenient. This work presents the first pilot experiments carried out at ILL-Grenoble with the high-efficiency Compton camera i-TED, hereby aiming at demonstrating its applicability for BNCT dosimetry by introducing real-time measurement of the boron concentration and imaging capabilities of spatial dose distribution. In this experiment, we measured the $^{10}$B uptake of different cancer cells of tongue squamous cell carcinoma, malignant melanoma and glioblastoma treated with BPA (80~ppm of $^{10}$B). The samples were irradiated with the thermal neutron spectrum of ILL-Grenoble and the 478keV $\gamma$-rays from the $^{7}$Li de-excitation after the neutron-boron reaction were registered both with the Compton imager and the high-sensitivity FIPPS HPGe array. These series of measurements allowed us to demonstrate the imaging capabilities of the Compton imaging device for the 478 keV $\gamma$-rays of interest for dosimetry in BNCT, as well as to assess its sensitivity, which was found to be below 1 $\mu$g of $^{10}$B. 

\end{abstract}

%% Keywords
\begin{keyword}
BNCT \sep dosimetry \sep Compton imaging \sep $^{10}$B concentration
%% keywords here, in the form: keyword \sep keyword

%% PACS codes here, in the form: 
%\PACS code \sep code
%% MSC codes here, in the form:
%%\MSC code \sep code

\end{keyword}

\end{frontmatter}

%% Add \usepackage{lineno} before \begin{document} and uncomment 
%% following line to enable line numbers
%% \linenumbers

%% main text
%%

%% Use \section commands to start a section
\section{Introduction} \label{Intro}
Boron neutron capture therapy (BNCT) is a promising cancer treatment technique, which exploits the high probability of a stable-boron nucleus ($^{10}$B) to capture thermal (25~meV) or epithermal neutrons (1 eV to 10 keV)~\cite{Sauerwein:12,Jin:22}. The cross section of $^{10}$B for absorbing neutrons is significantly greater than those of the elements in the human body, thus the accumulation of $^{10}$B in tumor cells along with the external irradiation of neutrons lead to high dose concentrations in the tumor. Additionally, the products of this reaction ($^7$Li nuclides and $\alpha$ particles) have high-linear energy transfer (LET) and high relative biological effectiveness (RBE). These characteristics make BNCT unique among external radiotherapies in its cellular-level selectivity~\cite{Malouff:21}. 

BNCT has been clinically investigated primarily for unresectable, locally advanced, and recurrent cancers, including glioblastoma multiforme, meningioma, head and neck cancers, lung cancers, and breast cancers~\cite{Malouff:21}. BNCT is undergoing significant expansion, marked by the establishment of new in-hospital, accelerator-based facilities worldwide~\cite{Kiyanagi:19}.  However, dosimetry in BNCT poses considerable challenges due to the indirect effect of neutrons with the elements present in the body and the unpredictable uptake of the boron compound 4‐borono‐L‐phenylalanine (BPA) used in clinical practice. As a result, treatment planning in BNCT is unconventional, relying primarily on estimates of boron uptake in tumors. These estimates are obtained through PET scans of patients using $^{18}$F-BPA, which provide an estimation of the tumor dose~\cite{Kabalka:97,Balcerzyk:20}. Additionally, the dose to organs at risk is estimated based on boron analytics in the blood~\cite{Koivunoro:19}.

\begin{figure}[!htbp]
\centering
\includegraphics[width=1.05\linewidth]{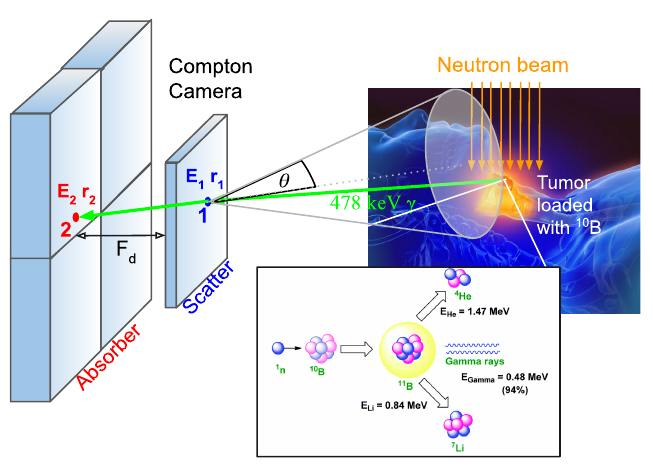}
\caption{Schematic view of Compton imaging applied to BNCT. In 93.7\% of the $^{10}$B(n,$\alpha$)$^{7}$Li reactions, a 0.478 MeV prompt gamma-ray is emitted which can be imaged with a Compon camera consisting of two position-sensitive detectors: Scatter and absorber.}
\label{fig:ComptonBNCT}
\end{figure}

A breakthrough in the field could be accomplished if one could measure --- in real time --- the boron concentration at the tumor and at organs of risk during treatment, as it is well known that the boron concentration varies with time. This would lead to an accurate dosimetry and more efficient treatments. A very interesting possibility comes from the detection, quantification and imaging of the prompt 478~keV $\gamma$-rays emitted in 94\% of neutron-capture reactions in $^{10}$B (see Fig.~\ref{fig:ComptonBNCT}). 

%In this context, intensive research has been carried out on imaging with Conventional (collimator based) gamma cameras or photo emission comptute tomography (SPECT)~\cite{Kobayashi:00,Rosenschoeld:06,Minsky:11,Yoon:14}. These systems rely on two-dimensional (2D) collimation of the gamma rays to produce images, which limits their sensitivity and spatial resolution. In particular, a substantial background is produced due to the interaction of neutrons with the collimator material and the secondary 511~keV radiation produced by pair production~\cite{Minsky:11}.  As an alternative, Compton imaging seems the most promising approach. The Compton camer localizes the position where the gamma ray is emitted by analyzing the kinematics of the Compton scattering. The Compton camera consists of at least two sensitive detectors, i.e., scatter detector and absorber detector, to score the position and energy loss of the Compton reaction, as shown in Fig.~\ref{fig:ComptonBNCT}. In recent years, several groups have explored the applicability of Compton imaging to BNCT on the basis of Monte Carlo simulations~\cite{Lee:15,Gong:17,Hou:22,Ramos-Lopez:23} but the experimental studies to date are still very scarce. In the work of Ref.~\cite{Sakai:23}, carried out with a commercial Si/CdTe Compton camera, a very limited efficiency of $\simeq$2$\cdot$10$^{-6}$ is reported, which may limit its applicability to real-time dose assessment in BNCT.

In this context, extensive research has been conducted on imaging using conventional gamma cameras (collimator-based) and single-photon emission computed tomography (SPECT) systems~\cite{Kobayashi:00,Rosenschoeld:06,Minsky:11,Yoon:14,Fatemi:19,Murata:21,Caracciolo:23}. These systems depend on two-dimensional (2D) collimation of $\gamma$-rays to produce images, which restricts the range of detectable energy (i.e., below 300 keV) and affects the attainable efficiency and spatial resolution~\cite{Gong:17}. Notably, significant background noise arises from the interaction of neutrons with the collimator material and the secondary 511 keV radiation generated by pair production~\cite{Minsky:11}. Another relevant source of background is the Compton continuum arising from the 2.2 MeV $\gamma$-rays from neutron capture in hydrogen. As a promising alternative, Compton imaging has been proposed. A Compton camera localizes the emission point of gamma rays by analyzing the kinematics of its Compton scattering in a multi-layer detector. This device comprises at least two sensitive detectors --- a scatter detector and an absorber detector --- that record the position and energy loss of the Compton reaction, as illustrated in Fig.~\ref{fig:ComptonBNCT}. In recent years, several research groups have investigated the potential of Compton imaging for BNCT using Monte Carlo simulations~\cite{Lee:15,Gong:17,Hou:22,Ramos-Lopez:23}. However, experimental studies remain limited. For instance, a study using a commercial Si/CdTe Compton camera~\cite{Sakai:23} reported a very low efficiency of $\simeq$2$\cdot$10$^{-6}$, which may constrain its applicability for real-time dose assessment in BNCT.

In order to provide further experimental evidence of the potential of Compton imaging for real-time determination of the boron distribution in BNCT, we present in this work the first pilot experiments carried out with i-TED, a high-efficiency Compton-based imager developed at Instituto de F\'isica Corpuscular (IFIC) within the HYMNS-ERC project~\cite{hymns}. The i-TED device and its Compton imaging performance are briefly introduced in Sec.~\ref{sec:ited}. Sec.~\ref{sec:Exp_ILL} describes the experiment carried out at ILL, in which several BPA and cell samples were irradiated with a thermal neutron spectrum and the 478keV $\gamma$-rays were registered both with the Compton imager and the high-sensitivity FIPPS HPGe array. The analysis of this pilot experiment, whose results are discussed in Sec.~\ref{sec:Results}, aims at studying the sensitivity to measure low $^{10}$B concentrations with i-TED compared to the a high-resolution low-background setup of FIPPS, and providing the first experimental insight on the ability and performance of the i-TED imaging system for the 478~keV $\gamma$-ray of interest for BNCT.

%% Labels are used to cross-reference an item using \ref command.

%% Use \subsection commands to start a subsection.
\section{Compton imaging with i-TED}\label{sec:ited}

\subsection{The i-TED Compton imager}
\begin{figure}[!htbp]
\centering
\includegraphics[width=0.8\linewidth]{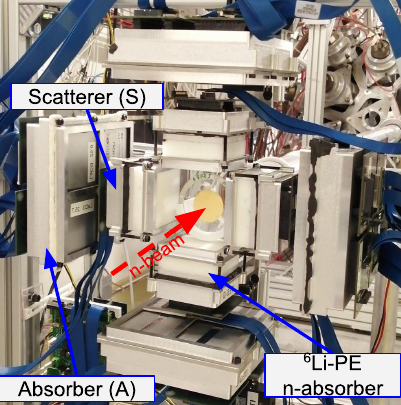}
\caption{Final i-TED array mounted in the CERN n\_TOF facility at CERN. The main components of on the four Compton cameras have been highlighted.}
\label{fig:iTED}
\end{figure}

In the pilot experiments presented in this work we evaluated the performance of one of the Compton cameras of i-TED, an array of four high-efficiency Compton cameras~\cite{Domingo:16,Babiano:21}. The original aim of this Compton camera was the enhancement of the sensitivity in high-resolution neutron TOF experiments of key stellar reactions~\cite{Domingo:16,Babiano:21,Lerendegui:23_NPA}.  After the development and characterization of the first demonstrator~\cite{Olleros:18,Babiano:19,Balibrea:21,Babiano:20}, the full array has been assembled and characterized in recent years~\cite{Gameiro:23} and used for experiments of astrophysical relevance at the CERN-n\_TOF-Facility~\cite{Lerendegui:23_NPA,Domingo:23}. A picture of the full i-TED Compton imaging array installed in the neutron beam of CERN-n\_TOF is shown in Fig.~\ref{fig:iTED}. Moreover, the applicability of the imaging capability of i-TED to range verification in Hadron Therapy by means of Prompt Gamma Imaging and Dual PET-Compton imaging has been explored with promising results~\cite{Lerendegui:22,Balibrea:22}, as well as its application for nuclear waste characterization~\cite{Babiano:24}. 

 Each of the i-TED Compton modules encompasses five position-sensitive detectors (PSDs) distributed in two parallel detection planes, Scatter (S) and Absorber (A), as shown in Figs.~\ref{fig:ComptonBNCT}-~\ref{fig:iTED}. The innovative i-TED design comprises a geometry where the four-fold size of the A-detector with respect to the S-detector enhances its efficiency, especially for low-energy $\gamma$-rays characterized by large scattering angles (see Fig.2 in Ref.~\cite{Babiano:20}). Each PSD contains a LaCl$_{3}$(Ce) monolithic crystal with a square-cuboid shape and a base surface of 50$\times$50~mm$^2$. 
 The LaCl$_{3}$ is hygroscopic and thus it is encapsulated in an aluminum housing. Each crystal base is coupled to a 2~mm thick quartz window, which is optically joined to a silicon photomultiplier (SiPM) from SensL (ArrayJ-60035-64P-PCB). The photosensor features 8$\times$8 pixels over a surface of 50$\times$50~mm$^2$. A 15~mm thick crystal is used for the PSD in the S-plane. Four 25~mm thick crystals are utilized for the PSDs placed in the A-plane (see Fig.~\ref{fig:iTED}). %In total, 1320 SiPM channels are biased and readout by means of front-end and processing PETsys TOFPET2 ASIC electronics~\cite{PETsys:16}, which allow reaching 300~kHz per individual detector. 
 In total, 320 SiPM channels per Compton module are biased and readout by means of front-end and processing PETsys TOFPET2 ASIC electronics~\cite{PETsys:16}, which allow acquiring event-rates of up to 500~kHz per individual detector. 
 In order to minimize gain shifts due to changes in the temperature of the experimental hall, every ASIC is thermally coupled to a refrigeration system composed by a Peltier cell, a heat-sink and a small-size fan (see Ref.~\cite{Babiano:20} for further details). More recently, a water-cooled system has been installed to further improve the temperature stability and, consequently, the energy resolution and the overall imaging performance.

An important feature for the application of i-TED to BNCT relies in its relatively low neutron sensitivity. To accomplish the latter goal LaCl$_3$(Ce) was preferred versus other options, owing to the relatively small integral capture cross section of Chlorine, and the small contribution of resonances in the keV-energy range of relevance. The Compton modules are supplemented with $^6$Li-loaded neutron-absorber pads of 20~mm thickness for further reducing the intrinsic neutron sensitivity of the array~\cite{Torres:24}. In addition, the successful implementation of ASIC-based TOF-PET readout electronics for Compton imaging has led to a rather compact and cost-effective system, when compared to other Compton imagers~\cite{Kataoka:2013,Nagao:2018}.

The original i-TED concept could be optimized for BNCT by reducing the crystal thicknesses~\cite{Torres:24}. Incidentally, this would result in a better position resolution and hence improves the final resolution of the reconstructed image~\cite{Balibrea:21}. An improved resolution, in particular in the depth of interaction, would have an especial impact in the attainable resolution with 3D tomographic imaging, which are highly desired for the final application to BNCT~\cite{Torres:24}.

\subsection{Compton imaging technique and performance}

%The Compton imaging technique is applied with i-TED using its high resolution position~\cite{Babiano:19,Balibrea:21} and energy~\cite{Olleros:18} sensitive radiation detectors arranged in two detection planes, so-called Scatter (S) and Absorber (A), operated in time coincidence. 
The Compton imaging technique is applied with i-TED using its high resolution position~\cite{Babiano:19,Balibrea:21} and energy~\cite{Olleros:18} sensitive radiation detectors operated in time coincidence. In order to reconstruct the Compton image, energy depositions (E$_{1}$ and E$_{2}$ in Fig.~\ref{fig:ComptonBNCT}) and 3D-localisation of the $\gamma$-ray hits (r$_{1}$ and r$_{2}$) in both layers are reconstructed for each coincidence event. Details on the reconstruction methods can be found elsewhere~\cite{Babiano:20,Babiano:21, Balibrea:22}. From the energies and positions, one can trace a cone, whose central axis corresponds to the straight line defined by the $\gamma$-ray interaction position in the two layers and its aperture $\theta$ is obtained from the measured energies using the Compton scattering formula, see Fig.~\ref{fig:ComptonBNCT}. The i-TED Compton modules embed the so-called dynamic electronic collimation technique~\cite{Babiano:20}, that allows one to remotely vary the distance between the A- and S-planes,F$_d$ in Fig.~\ref{fig:ComptonBNCT}, thereby optimizing the balance between efficiency and imaging resolution~\cite{Babiano:20}.

Most of the studies published to date on the Compton imaging performance of i-TED~\cite{Babiano:20,Babiano:21} have been based on an adaptation of the fast back-projection (BP) method of Ref.~\cite{Wilderman:98}. In order to boost the performance of i-TED for medical~\cite{Lerendegui:22,Balibrea:22} and industrial~\cite{Babiano:24} imaging applications, more evolved imaging algorithms have been recently implemented. This algorithms comprise the Stochastic Origin Ensemble (SOE) method~\cite{Andreyev:16}, the Maximum-Likelihood Expectation-Maximization (MLEM)~\cite{Wilderman:98b} and the Analytical Algorithm (AA) of Tomitani et al.~\cite{Tomitani:02}. More details on the implementation of the algorithms and a comparative study of their performance are given in Refs.~\cite{Lerendegui:22,Lerendegui:21_IEEE,Balibrea:22}. 

\begin{figure}[!t]
\centering
\includegraphics[width=0.8\linewidth]{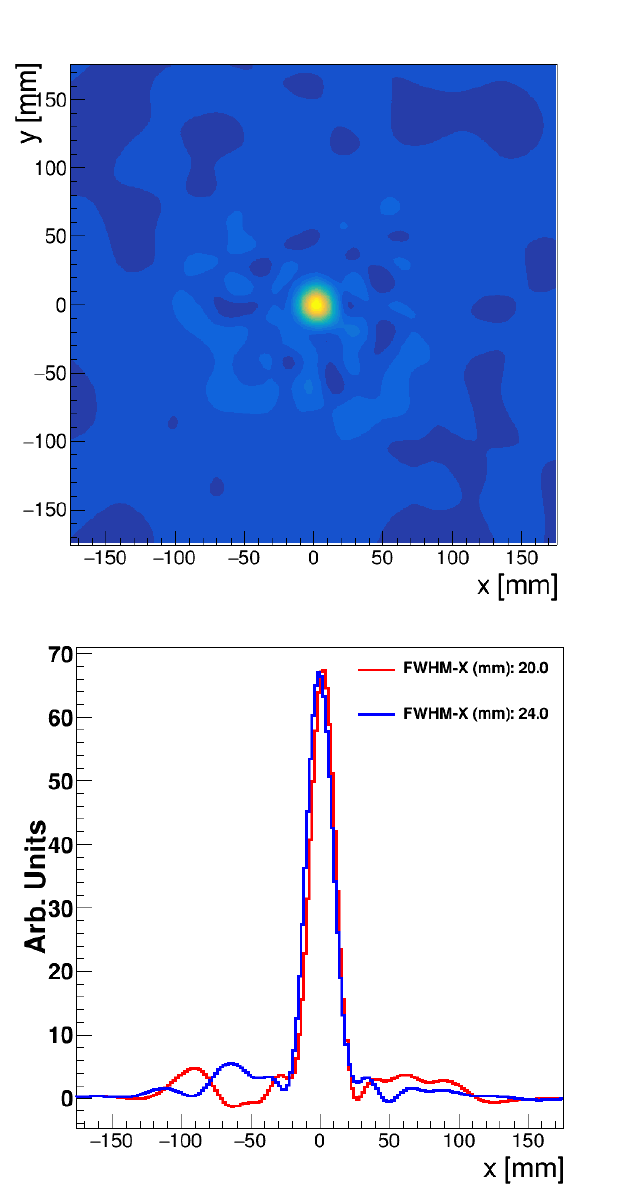}

\caption{Compton image reconstructed using the Analytical algorithm with one of the Compton cameras of i-TED for a $^{22}$Na point-like $\gamma$-ray source located at 10~cm from the scatter plane. The $\gamma$-ray of 1274~keV has been selected.}
\label{fig:ComptonLab}
\end{figure}

To illustrate the state-of-the-art imaging performance of i-TED, in Fig.~\ref{fig:ComptonLab} we present a Compton image reconstructed  using the best performing AA algorithm. To generate this image a $^{22}$Na source with an activity of 230~kBq was placed at 100~mm from one of the i-TED Compton cameras and measured for 900~s. The focal distance between the two PSD planes of the Compton camera (F$_{d}$ in Fig.~\ref{fig:ComptonBNCT}) was set to 50~mm, to optimize the angular resolution~\cite{Babiano:20}. Last, coincidence events corresponding to the 1274~keV $\gamma$-ray emitted in the decay of a $^{22}$Na source where selected. The bottom panel of the same figure shows the projection of the image onto the X and Y axis around the maximum. The projections show the accurate reproduction of the source coordinates (0,0) and allow one to quantify the spatial resolution, which is found to be 20~mm (FWHM), significantly better than the 40-50~mm obtained with an early prototype of i-TED at the same distance from the source using the back-projection algorithm~\cite{Babiano:21}. The spatial resolution determined from the projections in Fig.~\ref{fig:ComptonBNCT} corresponds to an angular resolution of about 5$^{\circ}$ to the level of one standard deviation ($\sigma$), clearly enhanced with respect to the early demonstrator of the current Compton cameras~\cite{Babiano:20}. %These results demonstrate that the recent upgrades of i-TED~\cite{Balibrea:21,Gameiro:23}, combined with the use of the more complex AA algorithm, have boosted the Compton imaging performance.

 These results demonstrate that the recent upgrades of i-TED~\cite{Balibrea:21,Gameiro:23}, combined with more advanced imaging-algorithms, have boosted the performance of the device to a level that could be of interest for medical applications. In particular, the first works aimed at range verification in hadrontherapy have already shown that i-TED could be sensitive to range variations of the prompt gamma ray profile of around 5~ mm~\cite{Lerendegui:22,Balibrea:22}. 

%The main drawback of the upgraded reconstruction algorithms discussed in this section is the computing time, which scales with the complexity of the involved models. 
At variance with mechanically-collimated systems, one of the main challenges to implement electronic (Compton) collimation for real-time dosimetry is the computing time required with advanced image-reconstruction algorithms, which scales with the complexity of the involved models.
According to our recent benchmark study~\cite{Lerendegui:22}, the fastest algorithm is BP followed by SOE and AA, being the latest about a factor 500 slower than BP. To overcome this limitation and really profit from its clearly superior performance, a GPU-boosted implementation of the AA algorithm based on the CUDA 11.1 toolkit was developed. The CUDA version of the Analytical Algorithm speeds-up the image reconstruction in a factor of ~120 with respect to the singled-threaded CPU version, and brings it back reconstruction times comparable to the SOE algorithm and only a factor ~5 slower than the fast back-projection. For the technical details on this work, the reader is referred to ~\cite{Lerendegui:22, Balibrea:22}.

\section{First pilot experiment at ILL} \label{sec:Exp_ILL}

\subsection{Neutron beam at ILL-FIPPS}
The experiment discussed herein was carried out at the FIPPS instrument, which exploits the thermal neutron beam of the H22 guide exiting the research reactor of Institut Laue-Langevin (ILL). This reactor produces the most intense continuous neutron flux in the world in the moderator region, 1.5$\cdot$10$^{15}$n/cm$^{2}$/s, with a thermal power of 58.3 MW.

\begin{figure}[!t]
\centering
\includegraphics[width=0.8\linewidth]{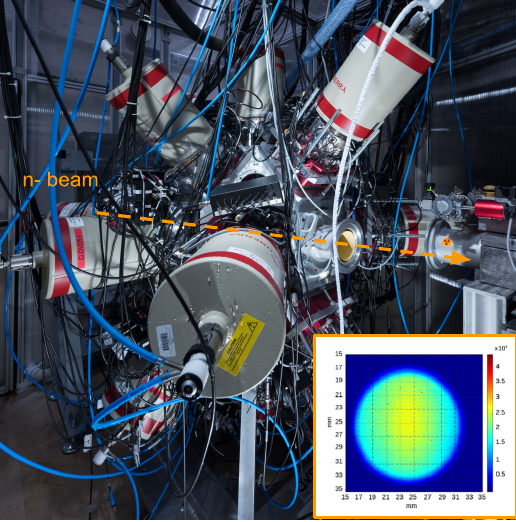}
\includegraphics[width=0.8\linewidth]{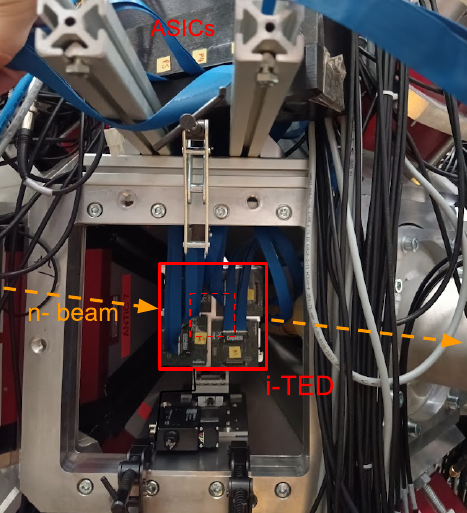}
\caption{Top: FIPPS instrument located in the H22 neutron guide. The insert shows the beam profile (from Ref~\cite{Michelagnoli:18}). Bottom: Experimental setup at ILL, where one of the FIPPS clover detectors was replaced by one of the Compton cameras of i-TED (bottom)}
\label{fig:FIPPS_Setup}
\end{figure}

The thermal neutron beam of the H22 guide is collimated in vacuum up to the target position, located at the center of the FIPPS array (see Fig.~\ref{fig:FIPPS_Setup}). The complex collimation system, which aims at reducing the beam-related $\gamma$-ray background in the detector, is composed of B$_4$C apertures followed by 5~cm lead absorbers and enriched $^6$LiF apertures. The latter are placed in the closest position to the FIPPS detectors~\cite{Michelagnoli:18}. The neutron beam at the target position, shown in the insert of Fig.~\ref{fig:FIPPS_Setup}, has a diameter of 1.5~cm and a flux of 5$\cdot$10$^{7}$n/cm$^{2}$/s~\cite{Michelagnoli:18}. 

\subsection{Experimental setup}
The FIPPS detector array, shown in Fig.~\ref{fig:FIPPS_Setup}, consists on eight germanium clover detectors arranged in the plane perpendicular to the neutron beam at the target position. Eight additional clover detectors with their anti-Compton (AC) shields from the IFIN collaboration are mounted at 45$^{\circ}$ with respect to this plane~\cite{Kandzia:20}. A Li-loaded tube mounted around the neutron beam is used to place the target in vacuum in the center of the setup and absorb scattered neutrons, thus minimizing the neutron-induced $\gamma$-ray background. FIPPS features a very high energy resolution of $\approx$2~keV at 1.3 MeV ($^{60}$Co) which, combined to the high efficiency of the 16 clovers ($\approx$8\% at 1.4 MeV and $\approx$18\% at 487~keV~\cite{Michelagnoli:18}) and the optimized background conditions, lead to a unique sensitivity to detect neutron-induced $\gamma$-rays, such as the 478~keV of the $^{10}$B(n,$\alpha$) reaction. However, it is worth noting that such a detection system would not be compatible with a clinical BNCT treatment room.

For this experiment, one of the Compton Cameras of i-TED was integrated in the FIPPS setup with the aim of comparing the sensitivity of the two setups. The bottom panel of Fig.~\ref{fig:FIPPS_Setup} shows the Compton imager installed in the gap left by one of the FIPPS clover detectors, that was taken out from the array. The Compton imager was vertically and horizontally aligned with the target and placed 150~mm away from its center at an angle of 45$^{\circ}$ with respect to the beam. The compactness of the device and electronics were key to adapt the setup to the limited space within the FIPPS array.

\subsection{Samples and measurements}

The campaign at ILL can be divided in two main experimental blocks, with different sample characteristics. A summary of all the samples, their main characteristics and irradiation times can be found in Table~\ref{tab:samples}. 

\begin{figure}[!htbp]
\centering
\includegraphics[width=\linewidth]{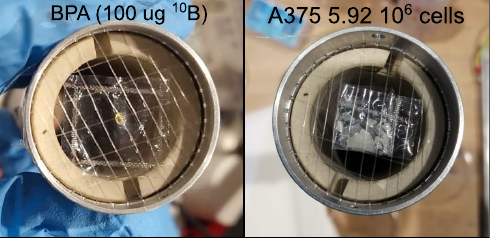}
\caption{Point-like sample of BPA containing 100$\mu$g of $^{10}$B (left) and sample of dried A375 cells (right) mounted on the sample holder of the FIPPS array.}
\label{fig:Samples}
\end{figure}

In the first part, a series of reference samples made out of a evaporated solution of fructose with BPA, with $^{10}$B equivalent contents spanning between roughly 300~ng and 500 $\mu$g, were irradiated in the thermal neutron beam. This allowed us to calibrate the detection response of each apparatus as a function of the $^{10}$B content, study their sensitivity and evaluate the Compton imaging performance of i-TED for a point-like sample of 487~keV $\gamma$-rays. The left panel of Fig.~\ref{fig:Samples} shows the sample containing 100~$\mu$g of $^{10}$B sealed in a Teflon bag and mounted on the target holder of FIPPS.  

After the previous characterization study, in the second block of the campaign a series of biological $^{10}$B-loaded samples were irradiated with thermal neutrons. %in FIPPS. 
Samples of tumor cell lines CAL33 (head and neck cancer), A172 (Glioblastoma) and A375 (melanoma) containing 4--18$\cdot$10$^{6}$ cells were treated with BPA (80 ppm of $^{10}$B in the medium) before drying them and depositing a monolayer of cells on a Teflon foil that is fully intercepted by the neutron beam (see right panel of Fig.~\ref{fig:Samples}). The irradiation time of each sample, listed in Table~\ref{tab:samples}, was estimated to achieve sufficient statistics with i-TED in S\&A coincidence --- $\approx$100~kcounts --- for the imaging of the 478~keV $\gamma$-rays.

\begin{table}[hb!]
\centering
\begin{tabular}{cccc}
\hline
Sample  & $^{10}$B mass ($\mu$g) & T$_{Irrad}$ (min)\\
 \hline
BPA + fructose & 500--0.6 & 5--900  \\
Fructose (bckg.) & 0   &  30 \\
\hline
Cell Line & Cells (x10$^{6}$)  & T$_{Irrad}$ (min)\\
\hline
CAL33 & 17.4 / 16.1 &  660 / 450 \\
A172 & 14.0 / 5.92 &  570 / 1080 \\
A375  & 18.4 / 11.2  & 435 / 360 \\
\hline

\end{tabular}
\caption{Properties of the samples used in this work and time of irradiation (T$_{Irrad}$).}
\label{tab:samples}       % Give a unique label
\end{table}

Besides the boron-loaded samples and cells, ancillary irradiations were carried out to assess the beam-related background with no BPA, by measuring a pure fructose sample deposited on the same Teflon bags. Moreover, the flux was measured by gold-foil activation. Last, ancillary calibration measurements with a $^{152}$Eu source were taken to calibrate in energy both the HPGe detectors of FIPPS and the five individual LaCl$_{3}$ detectors of the Compton imager.

\section{Results and discussion}\label{sec:Results}

\subsection{Sensitivity to $^{10}$B uptake}

\begin{figure}[!htbp]
\centering
\includegraphics[width=\linewidth]{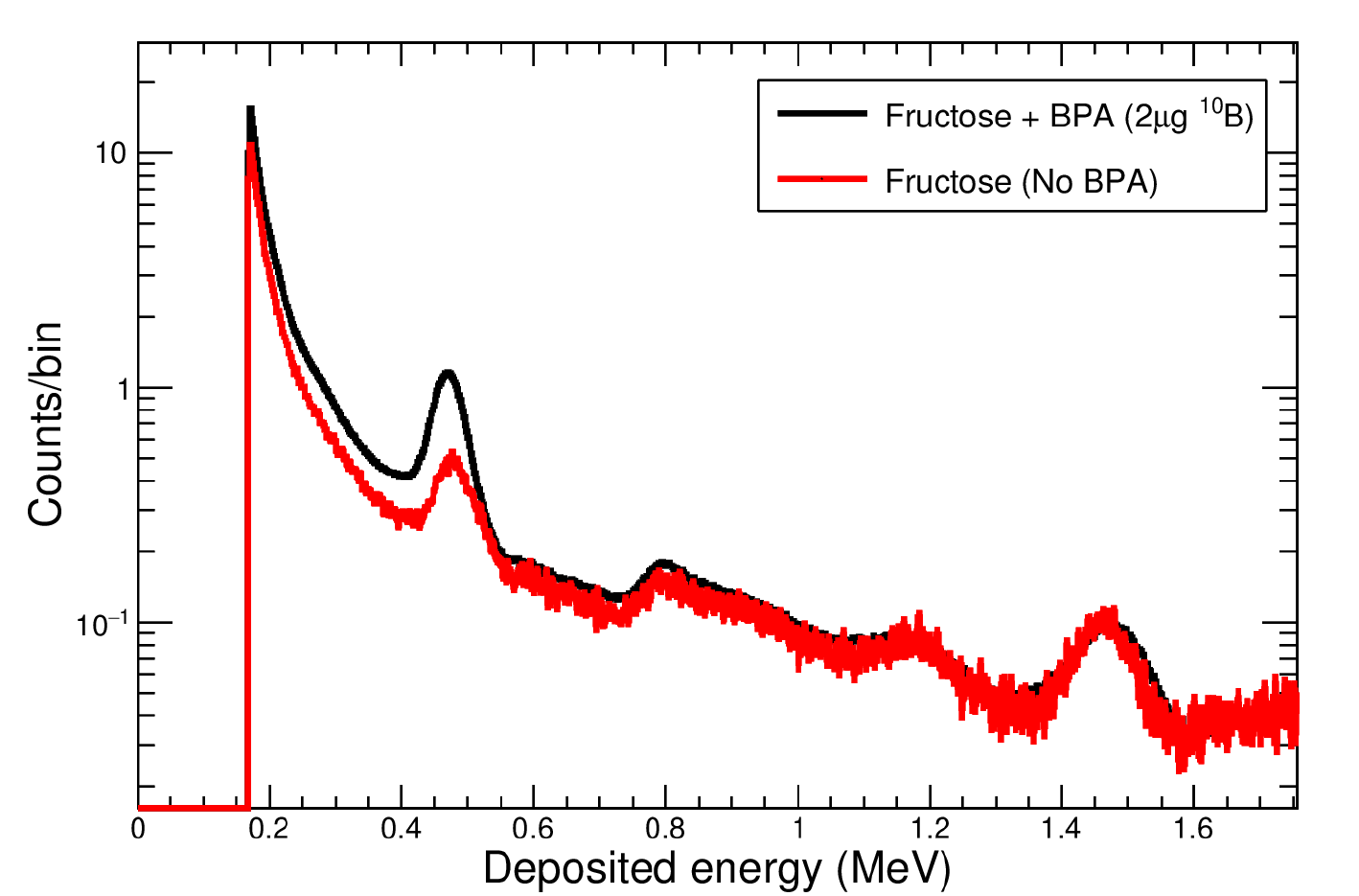}
\caption{Spectra of deposited energy for one of the absorber crystals of i-TED during the irradiation of the calibration sample containing 2$\mu$g of $^{10}$B compared to a background measurement.}
\label{fig:SpectraiTED}
\end{figure}

The aims of the sensitivity study were the determination of the $^{10}$B uptake in various cell lines treated with BPA and the comparison between the sensitivity of the high-resolution spectrometer FIPPS versus the medium-resolution Compton imager. Prior to this, we had to calibrate the detection response to the absolute $^{10}$B content of the samples and assess its linearity. To this end, we analyzed the spectra registered with both setups during the irradiation of BPA samples with known $^{10}$B concentration and studied the correlation between the 478~keV peak count rate and the $^{10}$B mass.

In the case of i-TED, the response of the individual detectors composing the Compton camera was individually calibrated to the $^{10}$B mass. To this aim, the energy spectra registered for each of irradiation were reconstructed applying the individual energy calibrations obtained from the measurement of a $^{152}$Eu sample and a temperature-dependent gain correction. Then, the count rate associated to the 478~keV line was determined from the excess of counts in each measurement with respect to the background counts. Fig.~\ref{fig:SpectraiTED} shows the deposited energy spectra of a detector of the absorber plane obtained in the irradiation of the BPA sample containing 2$\mu$g of $^{10}$B (black) compared to the measurement with no BPA content (red). To properly account for the underlying continuum, the 478~keV photo-peak was fitted in all the measurements to a Gaussian function and a linear background and the count rate were determined from the integral of the Gaussian fit carried out within the ROOT data analysis framework. The  uncertainty was derived from the uncertainties of the fit parameters.

\begin{figure}[!htbp]
\centering
\includegraphics[width=\linewidth]{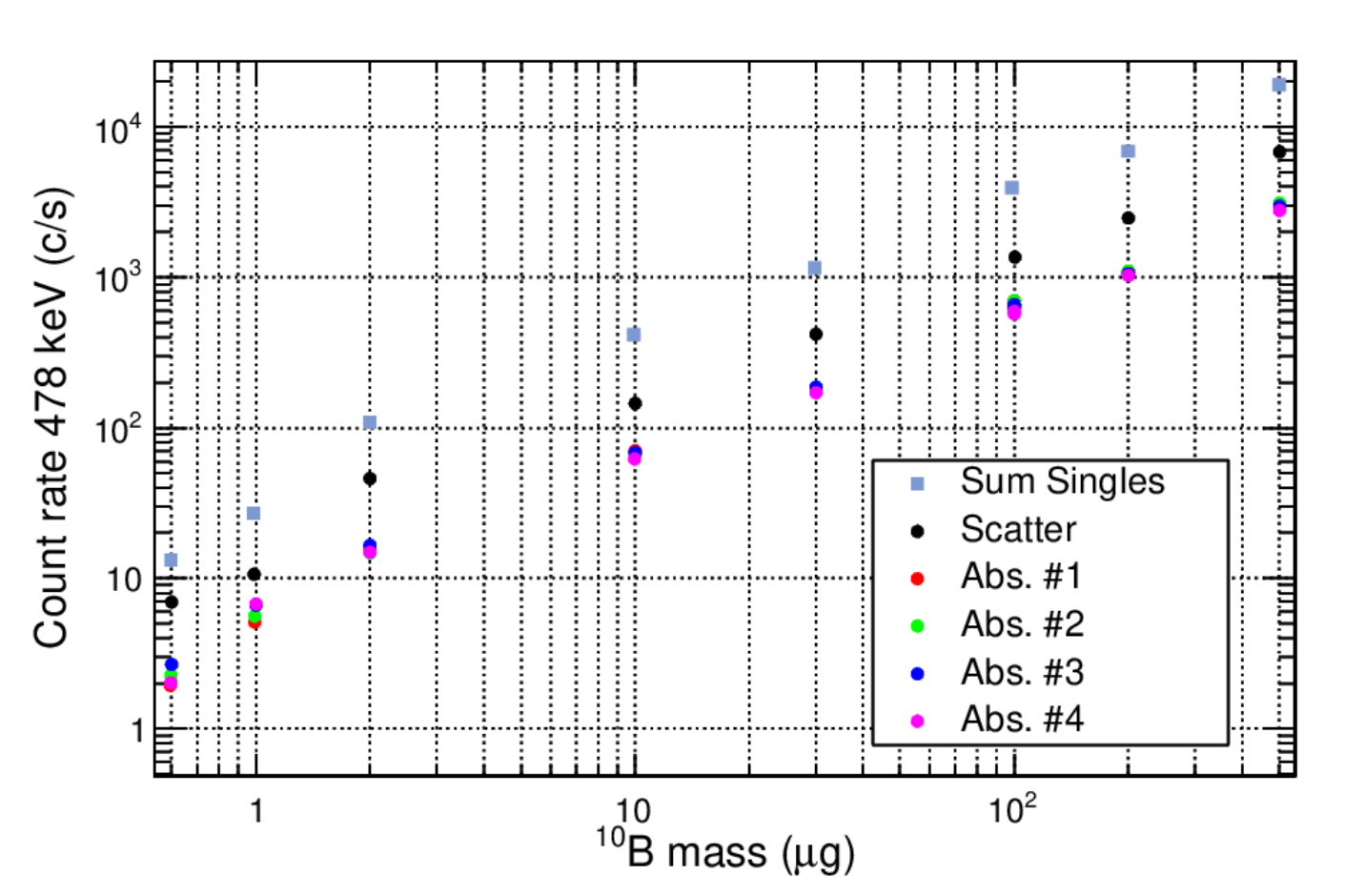}
\caption{Count rate in the 478~keV peak as a function of the $^{10}$B content of the BPA sample irradiated with neutrons. The different colors correspond to the single crystals of i-TED and the sum of all the crystals. Statistical uncertainties (2-3\% at maximum) are comparable or smaller than the points. 
}
\label{fig:iTEDB10Calibration}
\end{figure}

\begin{figure*}[!ht]
\centering
\includegraphics[width=0.7\linewidth]{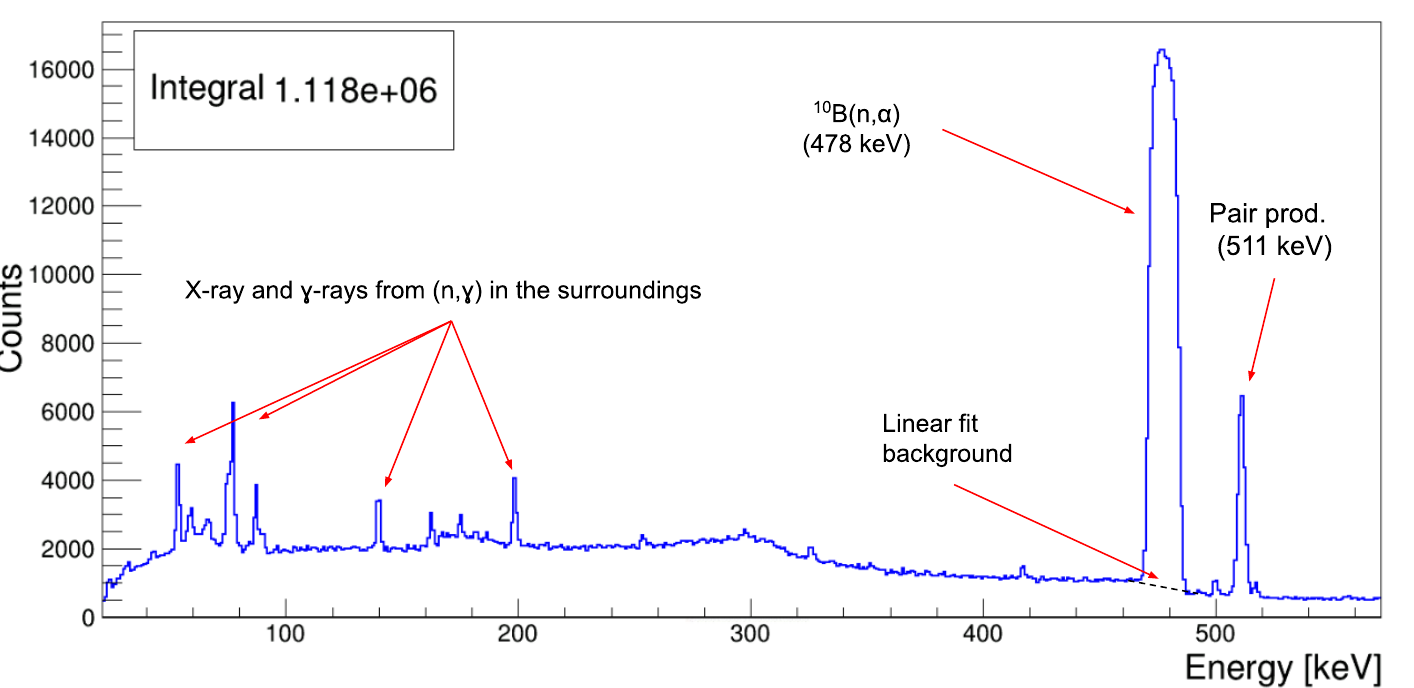}
\caption{FIPPS energy spectra acquired during the irradiation of the calibration sample with X$\mu$g of $^{10}$B. The main $\gamma$-ray lines below 511 keV are indicated}.
\label{fig:SpectraFIPPS}
\end{figure*}

The background-subtracted count rates resulting from the analysis of the different BPA samples irradiated with thermal neutrons are displayed in Fig.~\ref{fig:iTEDB10Calibration}. In this figure, the correlation between the 478~keV peak count rate and the $^{10}$B content of the irradiated samples is shown for the five individual crystals and the sum of all of them. Despite the large gamma-ray emission yields originated in the $^{10}$B(n,$\alpha$) reaction, expected to be $\simeq$10$^{6}$ gamma-rays per second for the 500$\mu$g sample, the count rate in single i-TED detectors does not exceed 10~kHz and, thus, no dead-time effects were noticed. Focusing in the low mass range (see Fig.~\ref{fig:iTEDB10Calibration}), the low background conditions within the FIPPS setup and the low neutron sensitivity of i-TED have pushed the sensitivity of i-TED down to at least 600~ng of $^{10}$B. 

As for FIPPS, the signals were processed with digital electronics (CAEN V1724 cards) with a sampling frequency of 100~MHz. The events were built in a 200~ns wide window from the data recorded in list-mode~\cite{Michelagnoli:18}. The data reduction chain merged the spectra of the full array of clover detectors and generated a energy-calibrated spectra, like the one displayed in Fig~\ref{fig:SpectraFIPPS}. The unique spectroscopic performance of the HPGe allows to perfectly separate the annihilation peak at 511 keV from the 478 keV peak. The latter has a broad rectangular line shape that is 8–10 times larger than that of normal $\gamma$-rays peaks as a result of Doppler broadening~\cite{Magara:98}. In order to build the calibration curve with FIPPS, the count-rate of the 478~keV photo-peak was determined after fitting the background with a linear fit (see Fig.~\ref{fig:SpectraFIPPS}) and subtracting its contribution. The resulting calibration showed a good linearity down to the lowest measured mass of 600~ng, as expected from the high sensitivity of FIPPS. On the contrary, samples of 100~$\mu$g and beyond, leading to count rates in the individual HPGe $\geq$10~KHz, were discarded from the calibration, due to the significant dead time losses. The latter limitation would represent a true additional drawback towards the implementation of HPGe detectors for dose monitoring in BNCT.

\begin{figure}[!htbp]
\centering
\includegraphics[width=1.0\linewidth]{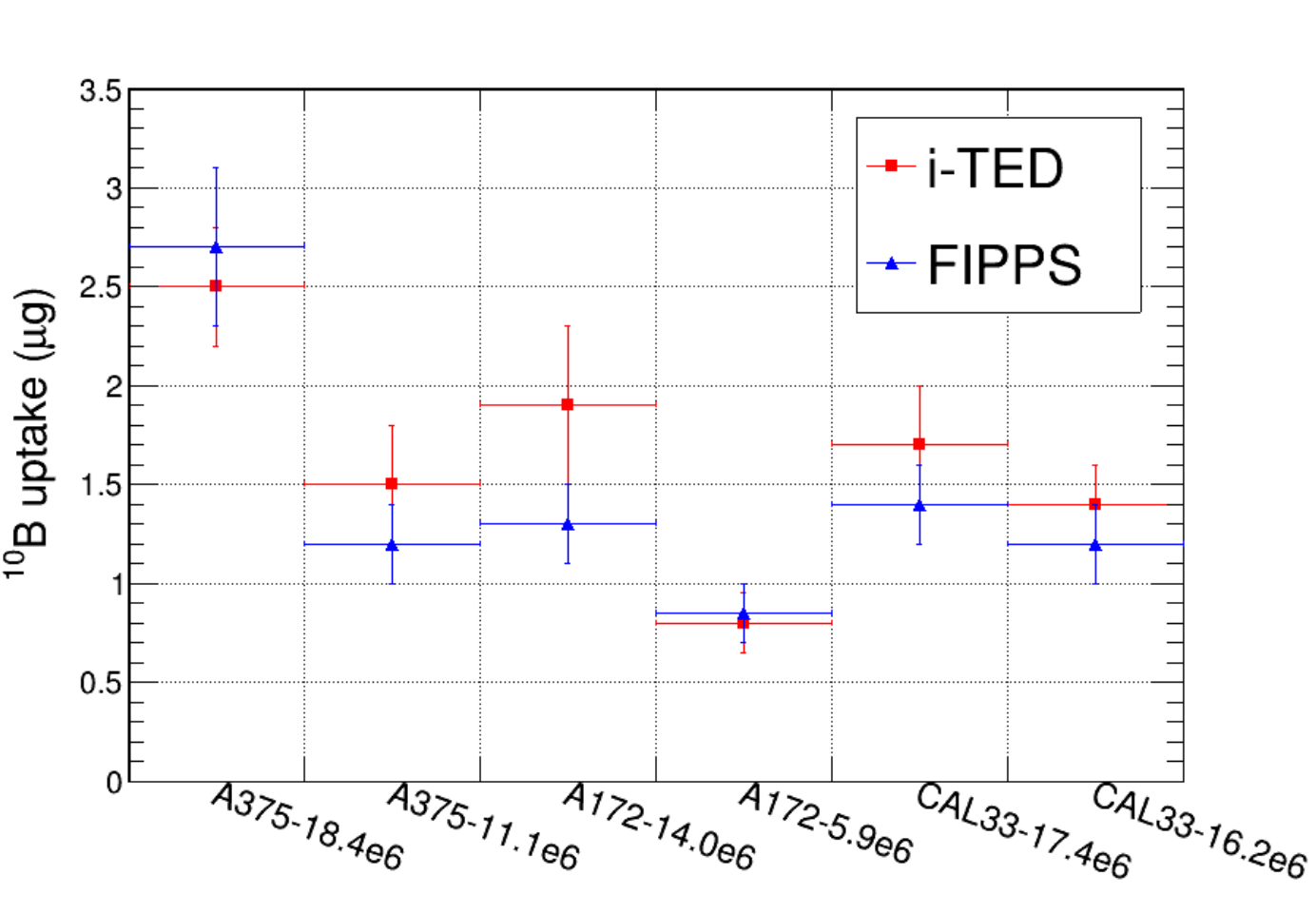}
\caption{$^{10}$B uptake of the different cell samples in this work determined from the measurements carried out with i-TED (red) and FIPPS (blue).}
\label{fig:B10uptake}
\end{figure}

Once the response of both setups had been calibrated, the $^{10}$B uptake of the different dried cell samples (see Table~\ref{tab:samples}) was determined from their irradiation in the thermal neutron beam of FIPPS. The background-subtracted counting rate of the 478~keV peak registered during the irradiation was transformed into $^{10}$B content, by means of a simple linear interpolation of the calibration curves. The resulting values for the $^{10}$B uptake of the six cell samples is presented in Fig.~\ref{fig:B10uptake}. This figure shows a consistent agreement --- within 1$\sigma$ for most cell samples, and below 1.5$\sigma$ for all of them--- for the results obtained with FIPPS and i-TED. The uncertainties in the results of Fig.~\ref{fig:B10uptake} include the systematic uncertainty in the $^{10}$B masses of the calibration samples, estimated to be 10~\%. In the case of i-TED, the final value for the $^{10}$B mass was computed as the average of the individual detectors, and the standard deviation of the individual results was incorporated as an additional systematic uncertainty. In summary, these results validate the capability of i-TED to perform absolute measurements of small boron concentrations in cells. Indeed, the results of $^{10}$B uptake of the cell lines in this work are being used in the pre-clinical cell survival study of Ref.~\cite{Alvarez:24}.

\subsection{Compton imaging in BNCT}

The second goal of the pilot experiment discussed herein was to address for the first time the potential of Compton imaging with i-TED in BNCT. To evaluate the imaging performance of this Compton imager, we profited from the same irradiations used for the sensitivity study.

\begin{figure}[!htbp]
\centering
\includegraphics[width=\linewidth]{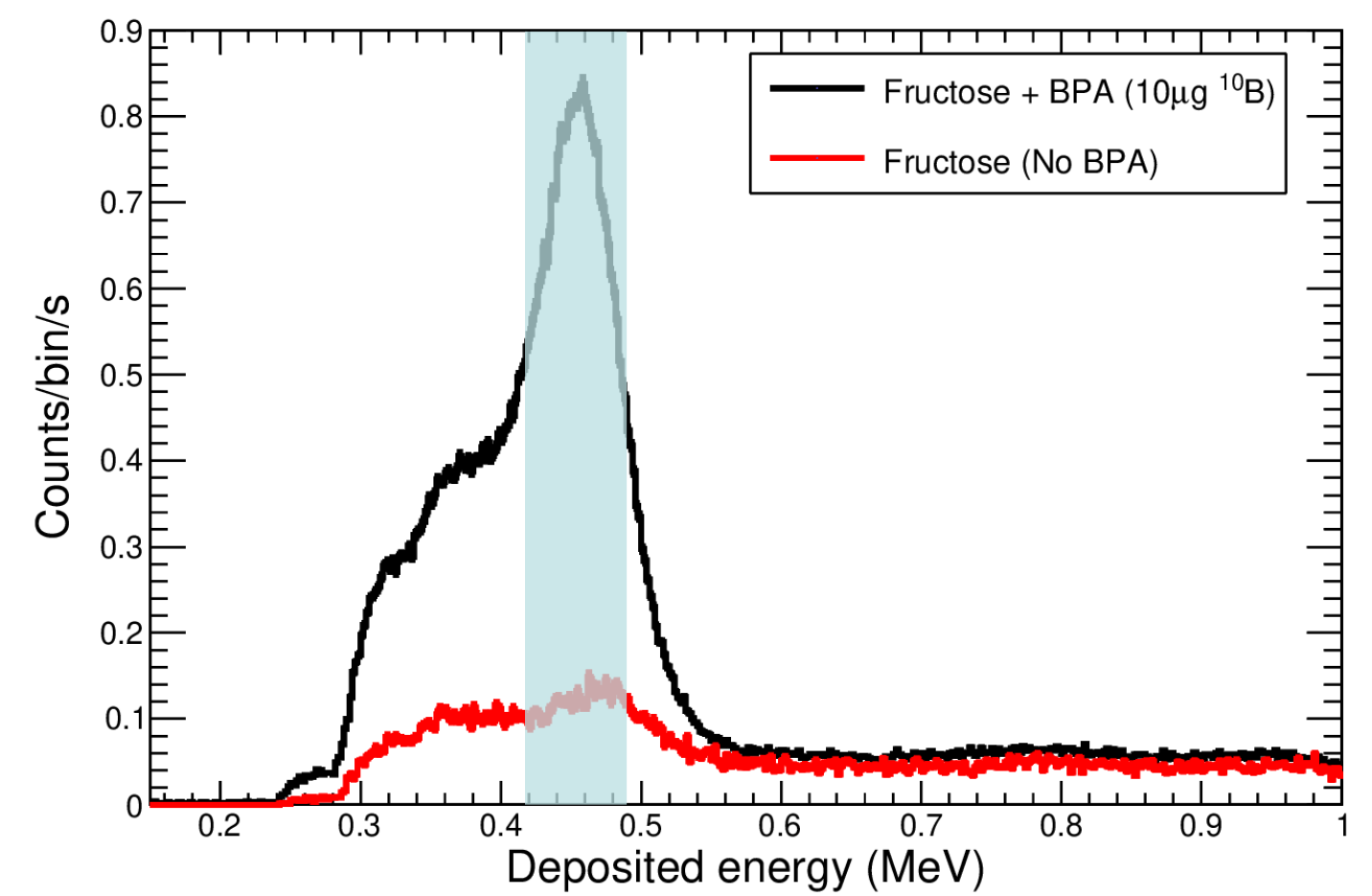}
\caption{Add-back deposited energy spectrum of i-TED during the irradiation of the BPA sample containing 10~$\mu$g of $^{10}$B (black) compared to the beam-related background counts, assesed with a fructose sample (red). The shadowed area indicates the energy selection around the 478~keV $\gamma$-ray line used for Compton imaging.}
\label{fig:ComptonAddback}
\end{figure}

In order to reconstruct the images, only events in time coincidence between the S- and A-layers were considered. This selection reduces the efficiency to about ~15\% relative to that of the Scatter detector~\cite{Babiano:21}. As introduced in Sec.~\ref{sec:ited}, the Compton imaging algorithms make use of the energy depositions and 3D-localisation of the $\gamma$-ray hits in both S- and A- position sensitive layers to compute the Compton cones and reconstruct the images. In order to image a single $\gamma$-ray energy, a selection is performed in the sum-energy value in time coincidence, referred to as add-back energy. As an example, the i-TED add-back spectrum for the BPA sample equivalent to 10~$\mu$g of $^{10}$B  is shown in Fig.~\ref{fig:ComptonAddback}, where it is also compared to a background run with no $^{10}$B. Signals below the 478~keV peak correspond to non-full-energy deposition events. 

The main experimental challenges for the Compton imaging of such low $\gamma$-ray energies arise from the poor energy resolution at low energy depositions and the efficiency and angular restrictions due to the low energy thresholds~\cite{Babiano:20}. As observed in Fig.~\ref{fig:ComptonAddback}, add-back deposited energies go down to 240-280 keV, associated to a low energy detection threshold of about 100 ~keV in the S- detector and 140-180~keV in the A-cristal. Moreover, the fair add-back energy resolution at this energy enables the clear distinction of the peak from the Compton continuum at lower add-back energies (see Fig.~\ref{fig:ComptonAddback}).

\begin{figure}[!t]
\centering
\includegraphics[width=0.8\linewidth]{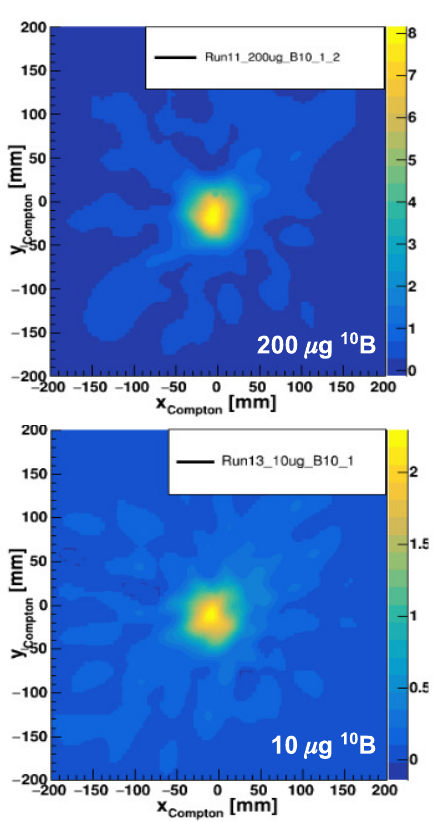}
\caption{Compton images reconstructed for the 478~keV $\gamma$-ray line emitted in the irradiation with neutrons of BPA samples containing 200$\mu$g (top) and 10$\mu$g of $^{10}$B (bottom).}
\label{fig:ComptonSamples}
\end{figure}

The coincident events falling inside the add-back energy window 0.42--0.5~MeV were selected to perform Compton imaging (see shadowed area in Fig.~\ref{fig:ComptonAddback}). Compton images were reconstructed on an image plane parallel to the detector planes and located in the sample position, 15~cm away from the Compton imager. Finally, the Compton images shown hereafter were obtained by implementing the Analytical Algorithm of Tomitani et al.~\cite{Tomitani:02}, introduced in Sec.~\ref{sec:ited}.

\begin{figure}[!htbp]
\centering
\includegraphics[width=0.8\linewidth]{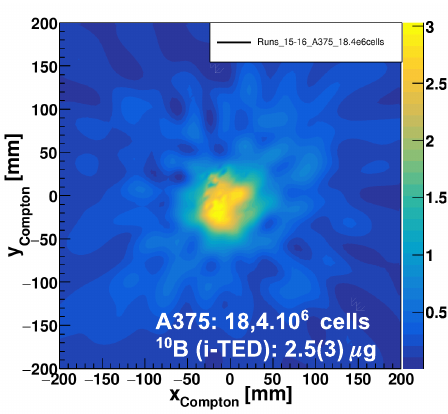}
\caption{Compton image reconstructed for the 478~keV $\gamma$-ray line emitted in the irradiation with neutrons of a sample containing 18.4$\cdot$10$^{6}$ A375 cells treated with BPA (80 ppm of $^{10}$B).}
\label{fig:ComptonCells}
\end{figure}

To provide a clearer insight on the ability and performance of the i-TED imaging system in a realistic situation with extended biological samples previously loaded with boron, 2D Compton images were also reconstructed for the different cell samples listed in Table~\ref{tab:samples}. The Compton image reconstructed from the irradiation of a dried monolayer of A375 melanoma cells is displayed in Fig.~\ref{fig:ComptonCells}. This sample, containing 18.4$\cdot$10$^6$ cells, was treated with 80~ppm BPA with an uptake of 2.5(3)~$\mu$g of $^{10}$B (see Fig.~\ref{fig:B10uptake}), and then irradiated in the FIPPS setup for 435~min.  The 2D image of Fig.~\ref{fig:ComptonCells} shows a relatively broad spatial resolution of 55–60~mm (FWHM) and less symmetric shape than the images of point-like BPA samples. This may reflect the uneven distribution of the pieces of dried cells in the neutron beam, as shown in Fig.~\ref{fig:Samples}. The image of a sample containing only 2.5$\mu$g of $^{10}$B, even though neutron induced background on Boron is present, shows the sensitivity of the device for the spatial localization of dose in BNCT. After this successful first pilot experiment, an optimization of the i-TED imaging system based on Monte Carlo simulations of a clinical scenario and more realistic tests have been undertaken~\cite{Torres:24}.

\section{Conclusions and outlook} \label{sec:Summary}
Online boron-uptake monitoring and imaging of the 478 keV line would be of great value for real-time dosimetry in BNCT. In this work, we have presented the first pilot experiment of the attainable sensitivity and imaging performance of i-TED, an array of four high-efficiency Compton cameras based on large LaCl$_{3}$ crystals, that is fully developed and has demonstrated its imaging capability in a wide range of fields. Its compactness, high count rate capability, low neutron sensitivity and GPU-Boosted algorithms make it particularly well suited for its application to real-time imaging in BNCT.

The experiment presented herein was carried out at ILL-Grenoble using the very high thermal neutron flux of 5$\cdot$10$^{7}$n/cm$^{2}$/s and using the FIPPS instrument as reference. One of the i-TED Compton imagers was embedded in the FIPPS HPGe array and both setups were run in parallel to detect the 478 keV $\gamma$-rays originated in $^{10}$B(n,$\alpha$)$^{7}$Li in BPA samples with known $^{10}$B content and tumoral cell samples (CAL33, A172, A375) treated with BPA (80 ppm $^{10}$B). The results show that, in the low-background conditions of the FIPPS setup, i-TED is sensitive enough to measure $^{10}$B concentrations $\leq$1~$\mu$g and is capable to determine absolute $^{10}$B concentrations in cell samples, obtaining consistent results to the high-sensitivity FIPPS array. These results make i-TED a promising non-destructive tool for pre-clinical in-vitro studies and open the door to future online dosimetry in clinical treatment. Last, this work represents the first experimental demonstration of the Compton imaging performance of i-TED for 478 keV $\gamma$-rays. 2D Compton images of cell samples of few $\mu$g have been reconstructed, despite the presence of boron-related background.

The successful pilot experiments described herein represent only a first proof-of-principle and are still far from the clinical conditions in BNCT because a very collimated beam of thermal neutrons with lower intensity than an actual BNCT facility was used. represent the first milestone towards the future applicability of large field-of-view Compton imaging in a more realistic BNCT preclinical scenario, where a broad neutron beam is irradiating a large biological sample loaded with $^{10}$B. The on-going work and future prospects, discussed in a recent publication~\cite{Torres:24}, include a new experimental campaign of the full i-TED array at ILL with more realistic phantoms, the optimization of the Compton camera in a Monte Carlo based clinical scenario, the development of 3D Compton algorithms, and, finally, future experimental campaigns with clinical beams. Last, besides the development of the Compton-based monitoring in BNCT, we aim at evaluating the performance in this field of a novel device which complements the $\gamma$-ray Compton technique with the imaging of the outgoing, mostly thermalized neutrons~\cite{Lerendegui:24,Patent}.

\section*{Declaration of competing interest}
The authors declare that they have no known competing financial interests or personal relationships that could have appeared to influence the work reported in this paper.
% https://www.elsevier.com/authors/journal-authors/policies-and-ethics/credit-author-statement

\section*{CRediT authorship contribution statement}
\textbf{J. Lerendegui-Marco:} Investigation, Methodology, Formal analysis, Data curation, Visualization, Writing - original draft.
\textbf{J. Balibrea-Correa:} Investigation, Methodology, Formal analysis, Data curation, Visualization, software.
\textbf{P.~\'Alvarez-Rodr\'iguez:} Investigation.
\textbf{V. Babiano-Su\'arez:} Investigation, Methodology, software.
\textbf{B.~Gameiro:} Investigation, Methodology, software. 
\textbf{I. Ladarescu:} Investigation, Software.
\textbf{C.~Méndez-Malagón:} Investigation.
\textbf{C. Michelagnoli}: Investigation, Resources. 
\textbf{I. Porras}: Investigation, Resources, Project administration, Funding acquisition, Writing - review \& editing.
\textbf{M.~Porras-Quesada}: Investigation.
\textbf{C. Ruiz-Ruiz}: Investigation, Resources, Writing - review \& editing.
\textbf{P. Torres-S\'anchez}: Investigation, Methodology, Writing - review \& editing.
\textbf{C. Domingo-Pardo:} Conceptualization, Investigation, Methodology, Supervision, Project administration, Funding acquisition, Writing - review \& editing.

\section*{Acknowledgment}
This work builds upon research conducted under the ERC Consolidator Grant project HYMNS (grant agreement No. 681740) and has been supported by the ERC Proof-of-Concept Grant project AMA (grant agreement No. 101137646). We also acknowledge funding from the Spanish Ministerio de Ciencia e Innovación under grants PID2022-138297NB-C21 and PID2019-104714GB-C21, as well as from CSIC under grant CSIC-2023-AEP128 and the Instituto de Salud Carlos III under grant DTS22/00147. We acknowledge support from the Severo Ochoa Grant CEX2023-001292-S funded by MCIU/AEI. Additionally, the authors thank the support provided by postdoctoral grants FJC2020-044688-I and ICJ220-045122-I, funded by MCIN/AEI/10.13039/501100011033 and the European Union NextGenerationEU/PRTR;  postdoctoral grant CIAPOS/2022/020 funded by the Generalitat Valenciana and the European Social Fund and a PhD grant PRE2023 from CSIC. Financial support from the Institut Laue-Langevin during the experimental campaign is also gratefully acknowledged.

%\bibliography{bibliography}

%\begin{thebibliography}{00

%\bibitem{lamport94}
%  Leslie Lamport,
%  \textit{\LaTeX: a document preparation system},
%Addison Wesley, Massachusetts,
%  2nd edition,
 % 1994.

%\end{thebibliography}

\end{document}